\documentclass[10pt]{article}

\usepackage[totalheight = 23cm, totalwidth = 17cm]{geometry}
\usepackage{amssymb,amsmath,amsfonts,amsbsy,graphicx,bm}
\usepackage{color}

\def\mathbi#1{\textbf{\em #1}}

\newcommand{\calH}{{\cal H}}

\newcommand{\calK}{{\cal K}}

\newcommand{\calR}{{\cal R}}

\newcommand{\CCC}{\mathbb{C}}
\newcommand{\CP}{\varphi}
\newcommand{\II}{\mathcal{R}}    
\newcommand{\SA}{c}
\newcommand{\SB}{d}
\newcommand{\SD}{c}
\newcommand{\SC}{e}

\newcommand{\QVEC}{\bm{Q}}
\newcommand{\up}[1]{{\rm #1}}

\newcommand{\beeq}{\begin{equation}}
\newcommand{\eneq}{\end{equation}}
\newcommand{\bear}{\begin{eqnarray}}
\newcommand{\enar}{\end{eqnarray}}
\newcommand{\nnn}{\nonumber \\}

\newcommand{\RA}{\rightarrow}

\newcommand{\HH}{\mathcal{H}}   
\newcommand{\gbar}{\bar g}      
\newcommand{\UU}{u}

\newcommand{\xvec}{\mathbi{x}}
\newcommand{\kvec}{\mathbi{k}}
\newcommand{\qvec}{\mathbi{q}}  
\newcommand{\vvec}{\mathbi{v}}

\newcommand{\mpc}{{\rm Mpc}}
\newcommand{\hmpci}{{h\mpc^{-1}}}

\begin{document}

\begin{titlepage}

\rightline{\footnotesize{APCTP-Pre2016-004}}

\begin{center}

\vskip 1.0 cm

{\LARGE \bf Exact analytic solution for non-linear density fluctuation
\\
in a $\Lambda$CDM universe
}

\vskip 1.0cm

{\large
Jaiyul Yoo$^{a,b}$ \, and \, Jinn-Ouk Gong$^{c,d}$	
}

\vskip 0.5cm

{\it
$^{a}$Center for Theoretical Astrophysics and Cosmology,
Institute for Computational Science\\
Universit\"at Z\"urich, CH-8057, Z\"urich, Switzerland
\\
$^{b}$Physics Institute, Universit\"at Z\"urich, CH-8057, Z\"urich, Switzerland
\\
$^{c}$Asia Pacific Center for Theoretical Physics, Pohang 37673, Korea
\\
$^{c}$Department of Physics, Postech, Pohang 37673, Korea
}

\vskip 1.2cm

\end{center}

\vspace{1.2cm}
\hrule \vspace{0.3cm}
\noindent {\sffamily \bfseries Abstract} \\[0.1cm]
We derive the exact 
third-order analytic solution of the matter density fluctuation in the 
proper-time hypersurface in a $\Lambda$CDM universe, accounting for the
explicit time-dependence and clarifying the relation to the initial condition.
Furthermore, we compare our analytic solution to the previous calculation
in the comoving gauge, and to the standard Newtonian perturbation
theory by providing Fourier kernels for the relativistic effects.
Our results provide an essential ingredient for a complete description
of galaxy bias in the relativistic context.
\vskip 10pt
\hrule
\vspace{0.6cm}
\end{titlepage}

\thispagestyle{empty}
\pagebreak

\noindent\hrulefill
{\linespread{0.75}\tableofcontents}
\noindent\hrulefill

\setcounter{page}{1}

\section{Introduction}
\label{sec:intro}
\setcounter{equation}{0}

The coming decades will witness the golden age in cosmology
with large scale galaxy surveys, as numerous
ambitious programs such as Euclid, WFIRST, LSST and so on,
will be in full operation, measuring
tens and hundreds of millions of galaxies in the sky and delivering an
unprecedented amount of data with unprecedented precision.
Taking full advantage of these impressive
experimental and observational developments
requires substantial advances in theoretical modeling.
In this regard, the recent development 
of the relativistic description of galaxy clustering \cite{YOFIZA09,YOO10}
calls for more endevour 
in theoretical description on large scales, where the relativistic effects in galaxy clustering become
important but has been ignored in the standard treatment of galaxy clustering
due to the lack of theoretical understanding and the large measurement
uncertainties.

The relativistic effects are intrinsically present in galaxy clustering, since all the
galaxy clustering observables are obtained by measuring light from the
source galaxies and the light propagation is subject to the same relativistic
effects that we measure in the cosmic microwave background (CMB).
One of the well-known relativistic effects 
in CMB is the Sachs-Wolfe effect, with which
photons lose their energy, climbing out of the gravitational potential
\cite{SAWO67}. The same relativistic effect
changes the observed redshift of galaxies
we measure in galaxy surveys, as it changes
the temperature of CMB photons. Another important example of the
relativistic effects in galaxy clustering is
the three-dimensional volume distortion in four-dimensional spacetime
mapped by the observed redshift and angular positions. 
A complete treatment of all the effects in galaxy clustering was given in
\cite{YOO09}, clarifying the effects that involve the intrinsic properties of the source galaxies (``source'' effects)
and those that involve the change of the volume the surveys cover (``volume'' effects).
Such relativistic
effects in galaxy clustering were previously unaccounted for in the standard
method. The full relativistic description of galaxy clustering was developed
in \cite{YOFIZA09,YOO10,YOHAET12}, providing new opportunities to probe
cosmology through subtle but unique relativistic effects:
see also \cite{BODU11,CHLE11,JESCHI12} for different derivations, and
see \cite{YOO14a} for review.

Going forward beyond linear theory, the
second-order relativistic description of galaxy clustering has been 
recently formulated
\cite{YOZA14,BEMACL14b,DIDUET14} to extract additional information from
the higher-order statistics such as the bispectrum. More work needs to be
done for the complete description, and in particular
one of the critical elements
in generalizing the formalism beyond linear order 
is galaxy bias that relates the galaxy number density to the underlying 
matter distribution. While it has been extensively studied in the
Newtonian framework, generalizing it to the relativistic framework requires
more work --- galaxy biasing was left out in \cite{DIDUET14}, and
the proper-time hypersurface was advocated in \cite{YOZA14,BEMACL14b}.

The proper-time hypersurface of nonrelativistic matter flows is a physically
well-defined 3-hypersurface a local observer can establish, who is moving together
with nonrelativistic matter flows such as dark matter or baryons on large
scales. This can be described by any choice of gauge conditions, but the
comoving gauge choice in a universe with a presureless medium allows
the global coordinate system to be aligned to the proper-time hypersurface,
facilitating the computation of the matter density fluctuation in the
proper-time hypersurface \cite{YOO14b}. This aspect is of particular importance 
when galaxy bias is considered. Beyond the linear order in 
perturbations, the spatial gauge conditions make a difference in physical
quantities, even with the same temporal gauge condition (comoving gauge 
in our case).
In \cite{BEMACL14b}, the synchronous comoving gauge was 
advocated for galaxy bias. However, it was shown \cite{YOO14b,HWNOET14} 
that the spatial coordinates in this gauge
condition trace the nonrelativistic matter flows and the direct computation
of the matter power spectrum in this coordinate is inadequate for galaxy
bias. In contrast, the comoving gauge condition with the spatial 
C-gauge condition fixes the spatial coordinates,
providing a natural framework to describe local dynamics in the relativistic
context \cite{YOO14b}. 
See Section~\ref{sec:non} for the detail of the gauge conditions mentioned above.

In this work, we derive the exact third-order analytic solution of the
matter density and the velocity fluctuations in a $\Lambda$CDM universe,
substantially extending the works
in the Einstein-de~Sitter (EdS) universe~\cite{Hwang:2012bi,YOGO15}.
The matter density fluctuation is the dominant contribution to galaxy 
clustering on all scales, and there is no relativistic correction to it at 
the linear order in perturbation. Compared to the leading-order power spectrum,
the subtle relativistic corrections that contain additional information require
the third-order relativistic calculation. Our work greatly expands the
calculations in \cite{Hwang:2012bi,YOGO15}, accounting for the explicit time-dependence
of each contribution, providing extensive studies of nonlinear relativistic
equations, and clarifying the difference in our solution to the
previous works \cite{HWNOET14,NOHW08,JEGOET11,Biern:2014zja}.

The organization of this article is as follows. In Section~\ref{sec:non} we derive
the nonlinear dynamical equations for the density and the velocity 
fluctuations in the proper-time hypersurface of nonrelativistic matter
flows. Full third-order analytic solutions are presented in 
Section~\ref{sec:solution}. Our analytic solutions are then compared to the
work in \cite{JEGOET11,Biern:2014zja} in Section~\ref{sec:comp}, 
and they are casted in terms of 
standard Fourier kernels in comparison to the standard Newtonian perturbation
theory in Section~\ref{sec:newt}. Finally, we end in Section~\ref{sec:dis}
with a discussion of further implication.
Throughout the article we will use $a,b,c,\cdots$ to represent the spacetime
indices and $i,j,k,\cdots$ to represent
the spatial indices. We assume a flat space
with the Friedmann-Robertson-Walker (FRW) metric.

\section{Nonlinear dynamical equations}
\label{sec:non}
\setcounter{equation}{0}

Here we briefly review the Arnowitt-Deser-Misner (ADM) formalism to describe the nonlinear dynamics 
and present our notation convention for the spacetime metric and its 
perturbations.

Given the spacetime metric~$g_{ab}$ and its coordinate system~$x^a$,
the ADM formalism considers spatial hypersurfaces labeled by its time
coordinate~$t$. The induced spatial metric~$h_{ij}=g_{ij}$ of these
3-hypersurfaces is treated as the dynamical degrees of freedom, subject to
the constraint equations.
The spacetime metric in the ADM formalism is conventionally written as
\cite{ADM,MTW} 
\beeq
\label{eq:ADM}
ds^2 = g_{ab}dx^adx^b = \left( -N^2+N^iN_i \right)dt^2+2N_idx^idt+h_{ij}dx^idx^j~,
\eneq
where the lapse function~$N$ represents the change in the proper time between
two spatial hypersurfaces with~$\Delta t$, the shift vector~$N^i$ 
represents the change in the normal direction~$n^a$ of the hypersurface,
and the spatial indices are lowered by the spatial metric~$h_{ij}$ (e.g., 
$N_i=h_{ij}N^j$). The normal vector (or often called the {\it normal observer}) to the 3-hypersurface 
is 
\beeq
\label{eq:four}
n^a=\left({1\over N},-{1\over N}N^i\right)~,
\qquad
n_a=g_{ab}n^b=(-N,~0)~,
\qquad 
n_an^a=-1~.
\eneq
Once the energy-momentum tensor~$T_{ab}$ is specified, 
the ADM fluid quantities can be derived, representing
the energy density, the momentum density, and the stress tensor
measured by the normal observer:
\begin{equation}
E = n_an_bT^{ab}=N^2T^{00}~,
\qquad
J_i = -n_aT^a_i=NT^0_i~,
\qquad
S_{ij} = T_{ij}~.
\end{equation}
In addition, the extrinsic curvature tensor $K_{ij}$ 
describes the local bending of 3-hypersurfaces
embedded in the four-dimensional spacetime:
\beeq
\label{eq:extrins}
K_{ij}={1\over2N}\left(N_{i:j}+ N_{j:i}-\dot h_{ij}\right)~,
\qquad 
K\equiv h^{ij}K_{ij}~,
\qquad
\bar K_{ij}\equiv K_{ij}- {1\over3}h_{ij} K~,
\eneq

We now make connections to a FRW universe, where the metric is described by
the usual FRW metric and small perturbations around the background.
Given the metric convention, the general relations between the ADM variables
and the metric perturbations can be derived \cite{NOHW04,HWNO07a}.
However, since we are interested in the proper-time hypersurface of
nonrelativistic matter flows, we first impose a gauge condition, greatly
simplifying the manipulation.

We choose the {\it comoving gauge} as our temporal gauge condition, in which
a local observer with the four velocity~$\UU^a$ moves along the flow of
nonrelativistic matter and sees vanishing
 energy flux in the observer rest frame. Furthermore,
the comoving gauge condition~$T^0{}_i=0$ is greatly simplified, if we
consider a universe composed of nonrelativistic matter only: as the
energy-momentum tensor in this case is $T_{ab}=\rho_m\UU_a\UU_b$
with the matter density~$\rho_m$, 
the comoving gauge condition becomes $\UU_i=0$, aligning the local 
observer~$\UU^a$ with the normal observer~$n^a$,
with which the ADM fluid description is directly applicable to the physical
system of interest.
The observer four velocity is then decomposed in terms of the 
shear~$\sigma_{ij}$ and the expansion~$\theta$ \cite{EHLER61,ELLIS71} as
\beeq
\label{eq:decomp}
\UU_{a;b}=n_{a;b}={1\over 3}\theta~h_{ab}+\sigma_{ab}~,
\qquad
\theta={n^a}_{;a}=-K~,\qquad
\sigma_{ij}=n_{(i;j)}=-\bar K_{ij}~,
\eneq
where the semicolon is the covariant derivative with respect to the spacetime
metric $g_{ab}$ and the induced metric~$h_{ab}=g_{ab}+\UU_a\UU_b$ 
is indeed the projection to the 3-hypersurface. The normal observer is
irrotational $\UU_{[a;b]}=0$, and the energy-momentum conservation of
the nonrelativistic matter flows imposes that the observer follows the
geodesic $a_a=\UU_{a;b}\UU^b=0$ and $N=1$ \cite{YOO14b}. 
Thus the coordinate time {\em exactly} corresponds to the proper time.

In addition, as our spatial gauge condition 
we choose the C-{\it gauge} \cite{NOHW04}, 
such that the spacetime metric takes the form
such that the off-diagonal term in the spatial metric $g_{ij}$ is removed:
\beeq
g_{00}=-1+N^iN_i\equiv -1-2~\alpha~,
\qquad
g_{0i}=N_i\equiv-\nabla_i\chi~,
\qquad
g_{ij}=h_{ij}\equiv a^2(1+2\CP)\gbar_{ij}~,
\eneq
where the spatial gradient~$\nabla^i$ is based on the background 
3-metric~$\gbar_{ij}$. It is noted \cite{NOHW04}
that the C-gauge condition leaves
no residual gauge freedom when combined with our temporal gauge condition.
We assume no vector or tensor perturbations in the spacetime metric.

In contrast,
as the spatial gauge condition one can opt to choose the B-$gauge$ \cite{NOHW04}, 
in addition to the same
temporal comoving gauge condition. This choice
is often called the {\it comoving-synchronous} gauge, and the metric becomes
\beeq
g_{00}=-1~,
\qquad 
g_{0i}\equiv N_i=0~,
\qquad 
g_{ij}=h_{ij}=a^2\left[(1+2\CP)\gbar_{ij}
+2\nabla_i\nabla_j\gamma\right]~.
\eneq
While this choice also corresponds to the proper-time 
hypersurface, the spatial coordinates are changing in time, 
tracing the nonrelativistic
matter flows in a way similar to the Lagrangian coordinates in the Newtonian
dynamics and leaving flows at rest in a given spatial coordinate.
In this work, no further investigation is made along this direction.

In our comoving C-gauge condition, the local observer~$\UU^a$ 
moving with the nonrelativistic matter flows becomes the normal observer~$n^a$,
facilitating the use of the ADM formalism in a physically meaningful way.
The ADM quantities in our case are greatly simplified as
\beeq
N=1~,
\qquad 
E=T^{00}=\rho_m~,
\qquad 
J_i=0~,
\qquad 
S_{ij}=\bar S_{ij}=S=0~,
\eneq
where the scalar part $S$ and the traceless part $\bar{S}_{ij}$ of the stress tensor $S_{ij}$ are
defined in the same manner as those of the extrinsic curvature $K_{ij}$ in~\eqref{eq:extrins}.

The relevant nonlinear equations based on the ADM variables are 
the conservation and constraint equations of energy and momentum,
and the trace and tracefree parts of the dynamical equations. 
The complete set of the ADM equations can be found in 
\cite{ADM,BARDE80,BARDE88} and we do not present them here.
At the background level in perturbations, the nonlinear equations correspond to
the familiar matter density conservation and the Friedmann equations.
With the background evolution removed, the nonlinear dynamical equations
yield a series of nonlinear perturbation equations to be solved
for the density fluctuation~$\delta \equiv \rho_m/\bar\rho_m-1$, with $\bar\rho_m$
being the background density, and the perturbation in the
extrinsic curvature~$\kappa \equiv 3H+K$.
The master dynamical equations are the conservation equation
and the Raychaudhuri equation:
\begin{align}
\label{eq:mast1}
\dot\delta-\kappa&= N^i\nabla_i\delta+\delta\kappa~,
\\
\label{eq:mast2}
\dot\kappa+2H\kappa-4\pi G\bar\rho_m\delta&=
N^i\nabla_i\kappa+{1\over3}\kappa^2+\sigma^{ij}\sigma_{ij}~,
\end{align}
supplemented by the constraint equations:
\begin{align}
\label{eq:enecons}
\delta R&= \sigma^{ij}\sigma_{ij}+4H\kappa-{2\over3}\kappa^2+16\pi G\bar
\rho_m\delta~,
\\
\label{eq:constpert}
\frac23\nabla_i\kappa&=\sigma^j{}_{i:j}~.
\end{align}

In order to solve these dynamical equations perturbatively, 
we need to compute the nonlinear perturbation variables at each order
such as the metric perturbations and the geometric quantities of
3-hypersurfaces. For example, up to the third order in perturbations,
$N^i$, $\kappa$ and $\sigma_{ij}$ can be written by using
\eqref{eq:extrins} and~\eqref{eq:decomp} as
\begin{align}
\label{eq:ADMNi}
N^i&= -{1\over a^2}\nabla^i\chi\left(1-2\varphi+4\varphi^2\right)~,
\\
\label{eq:kappa}
\kappa&= -3\dot\CP-{\Delta\over a^2}\chi
+6\CP\dot\CP+{1\over a^2}\left[2\CP\Delta\chi(1-2\CP)
-\nabla_i\chi\nabla^i\CP (1-4\CP)\right]~,
\\
\label{eq:shear}
\sigma_{ij}&= \left(\nabla_i\nabla_j-\frac13\delta_{ij}\Delta\right)\chi
-2\left(\nabla_{(i}\CP\nabla_{j)}\chi-\frac13\delta_{ij}\nabla^k\CP
\nabla_k\chi\right)(1-2\CP)~.
\end{align}
The momentum constraint equation can be arranged as
\begin{align}
\label{eq:momentum}
\kappa+{1\over a^2}\Delta\chi&=
{1\over a^2}\left[2\CP\Delta\chi(1-2\CP)
-\nabla^i\CP\nabla_i\chi(1-4\CP)\right]
\nnn
& \quad
+{3\over2a^2}\Delta^{-1}\nabla^i\bigg[
\nabla_i\chi\Delta\CP+\nabla_j\nabla_i\CP\nabla^j\chi
-4\CP\left(\nabla^j\chi\nabla_j\nabla_i \CP
+\Delta\CP\nabla_i\chi\right)-\left(\nabla_i\chi\nabla_j\CP
+3\nabla_j\chi\nabla_i\CP\right)\nabla^j\CP\bigg]~.
\end{align}
We ignored the vector and the tensor contributions (but see
\cite{HWJENO15}).
Furthermore, combining the definition of~$\kappa$ in~\eqref{eq:kappa} with the
ADM momentum constraint, we derive the dynamical equation for the
curvature potential:
\beeq
\label{eq:time}
\dot\varphi = 2\CP\dot\CP-{1\over2a^2}\Delta^{-1}\nabla^i\bigg[\nabla^j\chi
\nabla_j\nabla_i\CP+\Delta\CP\nabla_i\chi-4\CP\left(\nabla^j\chi\nabla_j
\nabla_i \CP+\Delta\CP\nabla_i\chi\right)-\left(\nabla_i\chi\nabla_j\CP
+3\nabla_j\chi\nabla_i\CP \right)\nabla^j\CP\bigg]~.
\eneq
At the linear order in perturbations, the curvature potential is a
time-independent spatial function set by the 
initial condition~$\CP^{(1)}\equiv\II(\xvec)$. This remains true
to all orders in perturbation on super-horizon scales, where the gradient
terms are negligible. On sub-horizon scales, the curvature potential evolves
in time beyond the linear order in perturbations, and we need to
evaluate the time-dependence of the nonlinear terms in~\eqref{eq:time} 
before we integrate to obtain the time-evolution of the curvature potential.

\section{Analytic solutions of the matter density fluctuation}
\label{sec:solution}
\setcounter{equation}{0}

Armed with the nonlinear equations in Section~\ref{sec:non}, 
in this section we now derive the
third-order analytic solution of the matter density fluctuation~$\delta$ in a $\Lambda$CDM
universe.

\subsection{Battle plan}
\label{ssec:plan}

The analytic derivation of the third-order solutions in general relativity
inevitably involves many steps technical and lengthy in nature, so we
start by presenting the master differential equation and the overall strategy
to solve the differential equation at each order in perturbations.

Using the continuity equation~\eqref{eq:mast1},
the ADM energy constraint~\eqref{eq:enecons} can be rearranged as the
master differential equation for~$\delta$:
\beeq
\label{eq:main}
\HH\delta'+\frac32\HH^2\Omega_m\delta={a^2\over4}\left(
\delta R-\sigma^{ij}\sigma_{ij}+{2\over3}\kappa^2
+4HN^i\nabla_i\delta+4H\delta\kappa\right)~,
\eneq
where the prime is the derivative with respect to the conformal time
$d\eta=dt/a$, $\HH=a'/a=aH$ is the conformal Hubble parameter 
and $\Omega_m = 8\pi G\bar\rho_m/(3H^2)$. 
The left-hand side (LHS) of \eqref{eq:main} is linear in~$\delta$,
and the right-hand side (RHS) is composed of at least quadratic terms,
except the intrinsic curvature of 3-hypersurface  $\delta{R}$, such that
$n$-th order solution~$\delta^{(n)}$ can be used to compute $(n+1)$-th order terms in RHS$^{(n+1)}$ and 
derive $(n+1)$-th order solution~$\delta^{(n+1)}$ in the LHS.
Once $n$-th order solution~$\delta^{(n)}$ is derived,
the solution~$\kappa^{(n)}$ can be obtained algebraically by using the ADM energy 
constraint, explicitly written as
\begin{align}
\label{eq:ADMene}
\frac32H^2\Omega_m\delta+H\kappa+{1\over a^2}\Delta\CP&= {1\over6}\kappa^2
+{1\over 12a^4}\left[(\Delta\chi)^2-3\nabla_i\nabla_j\chi\nabla^i
\nabla^j\chi\right](1-4\CP)+{1\over a^2}\left(4\varphi\Delta\varphi+\frac32
\nabla^i\varphi\nabla_i\varphi\right)  
\nnn
& \quad 
+{1\over a^4}\left(\nabla^j\nabla^i\chi\nabla_j\CP\nabla_i\chi
-{1\over3}\nabla^i\CP\nabla_i\chi\Delta\chi\right)
-{3\over a^2}\CP\left(3\nabla^i\CP\nabla_i\CP+4\CP\Delta\CP\right)~.
\end{align}

The homogeneous solution that satisfies \eqref{eq:main} with vanishing
RHS is readily derived as
$\delta_h\propto H$ and is identified as the usual decaying mode in the standard
Newtonian solution, which we ignore henceforth. The particular solution
with nonvanishing RHS corresponds to the growing mode solution:
\beeq
\delta_p = \delta_h\int{ d\tau\over\delta_h}\left({\up{RHS}\over\HH}\right)
=H\int  dt\left({\up{RHS}\over\HH^2}\right)~,
\eneq
where RHS is of dimension two. To compute RHS of \eqref{eq:main} at each order in perturbations,
we split RHS as the sum of perturbative expansion terms:
\beeq
\label{eq:rhssep}
\up{RHS}\equiv \sum_{n} \text{RHS}^{(n)} = \up{RHS}^{(1)} + \up{RHS}^{(2)} + \up{RHS}^{(3)} + \cdots~,
\eneq
where the superscripts represent the order of each term in perturbative
expansions. First, we write RHS$^{(n)}$ at each perturbation order
as the sum of terms RHS$_m^{(n)}$:
\beeq
\up{RHS}^{(n)}(t,\xvec) \equiv  \sum_{m=1}^n 
\up{RHS}_m^{(n)}(t,\xvec)~,
\eneq
where the subscript~$m$ indicates that the time-dependence 
in the EdS universe scales as $\up{RHS}^{(n)}_m(t,\xvec)\propto D_1^m(t)$.
Each of these RHS$_m^{(n)}$ is decomposed as the sum of the 
scale-dependent and time-dependent functions:
\beeq
\up{RHS}^{(n)}_m(t,\xvec)= \!\!\!\!\!
\sum_{I=A,B,\cdots} \!\!\!\!\! X_{mI}^{(n)}(\xvec)T_{mI}(t)~,
\eneq
where the subscript $I$ denotes different time dependences that
become identical as~$T_{mI}(t)\propto D_1^m(t)$ in
the EdS universe, and $X_{mI}^{(n)}(\xvec)$ is a time-independent
but scale-dependent function at $n$-th order in perturbations.

According to this decomposition, the growing mode solution will be
the sum of individual solutions~$\delta^{(n)}_{mI}$ with
corresponding RHS$_{mI}^{(n)}$:
\beeq
\delta^{(n)}_{mI}(t,\xvec)=H\int dt\left({\up{RHS}_{mI}^{(n)}\over\HH^2}\right)
=D_{mI}(t) X_{mI}^{(n)}(\xvec)
\qquad \text{with} \qquad
D_{mI}(t)  = H \int dt ~\frac{T_{mI}(t)}{\HH^2}~,
\eneq
where $D_{mI}(t)$ is of dimension minus two. It is noted that 
$\delta_{mI}^{(n)}$ is at $n$-th order in perturbations and its 
time-dependence~$D_{mI}(t)$
is determined by the time-dependent function $T_{mI}(t)$ in RHS$_{mI}^{(n)}$.
Therefore, the full solution is then
\begin{equation}
\label{eq:growth}
\delta_p=\delta^{(1)}+\delta^{(2)}+\delta^{(3)}+\cdots
\qquad \text{where} \qquad
\delta^{(n)}(t,\xvec) = \sum_{m,I} \delta_{mI}^{(n)}(t,\xvec) \, .
\end{equation}
The main result of this section is this analytic solution up to third order, 
given by \eqref{eq:sol-linear}, \eqref{eq:sol-2nd} and \eqref{eq:sol-3rd}.

A further manipulation can be made to facilitate the computation by defining
the logarithmic growth rate~$f_{mI}(t)$ associated with~$D_{mI}(t)$:
\beeq
f_{mI}(t) \equiv {d\ln D_{mI}(t)\over d\ln a}~,
\qquad \qquad
D_{mI}^{~\prime} = \HH f_{mI}D_{mI}~.
\eneq
Using the logarithmic growth rate, the LHS of~\eqref{eq:main} can be
written as
\beeq
\up{LHS}
\left[\delta_{mI}^{(i)}\right] = \HH{\delta_{mI}^{(i)}}' + {3\over2}\HH^2\Omega_m\delta_{mI}^{(i)} 
\equiv \HH^2f_{mI}\Sigma_{mI}\delta_{mI}^{(i)}~,
\qquad \qquad
\Sigma_{mI}(t)= 1+{3\over2}{\Omega_m\over f_{mI}}~,
\eneq
and the growth rate is then related to the logarithmic growth rate as
\beeq
\label{eq:loga}
D_{mI}(t)={T_{mI}\over\HH^2f_{mI}\Sigma_{mI}}~,
\qquad\qquad
f_{mI}(t)={{D_{mI}'}\over \HH D_{mI}}=-{3\over2}\Omega_m+{T_{mI}\over
\HH^2D_{mI}}={T_{mI}\over\HH^2\Sigma_{mI}D_{mI}}~,
\eneq
providing a convenient way of computing the logarithmic growth rate
without taking numerical differentiation of the growth factor.

\subsection{Linear- and second-order solutions}
\label{ssec:known-sol}

We start by deriving the well-known linear- and second-order solutions to provide
the guidance of the strategy laid in Section~\ref{ssec:plan}.
At the linear order in perturbations, the RHS of~\eqref{eq:main} is simply
\beeq
\up{RHS}^{(1)}(\xvec)
= -\Delta\varphi^{(1)}(\xvec)
\equiv -\Delta\II(\xvec)=X_1^{(1)}(\xvec)
\qquad \text{with} \qquad 
T_1(t)=1~,
\eneq
so that the linear-order growth solution is then
\begin{align}
\label{eq:D1}
D_1(t) & = H\int {dt\over\HH^2}={1\over\HH^2f_1\Sigma_1}~,
\\
\label{eq:sol-linear}
\delta_1^{(1)}(t,\xvec) & = D_1(t)X_1^{(1)}(\xvec) = -{\Delta\II(\xvec)\over\HH^2f_1\Sigma_1}~,
\end{align}
where the linear-order growth factor~$D_1$ needs to be numerically integrated
before the logarithmic growth rate~$f_1$ is obtained.\footnote{In a 
$\Lambda$CDM universe, the linear-order growth factor can be analytically
computed in terms of the associated Legendre function of the second kind
\cite{WEINB87}, while it still needs to be 
numerically evaluated.} The linear growth factor~$D_1$ is identical to
one in the standard Newtonian description, once normalized to remove its
dimension at some epoch (see Section~\ref{ssec:spt}).
According to the ADM energy constraint~\eqref{eq:ADMene} and 
the ADM momentum constraint~\eqref{eq:momentum}, we can derive
the linear-order perturbation to the extrinsic curvature 
$\kappa_1^{(1)}$\footnote{
Inspecting the time-dependence of the continuity equation~\eqref{eq:mast1},
we find this relation remains valid to all orders in perturbation, i.e.
for the constant terms on the RHS of \eqref{eq:main} irrespective of perturbation order,
\begin{equation*}
\label{eq:full1}
{\kappa_1(t,\xvec)\over Hf_1}=\delta_1(t,\xvec)={\up{RHS}_1(\xvec)\over\HH^2
f_1\Sigma_1}~,
\end{equation*}
with $\text{RHS}_1 = \sum_n \text{RHS}_1^{(n)}$ [see \eqref{eq:2nd-D1} and \eqref{eq:3rd-D1}].
}
and the scalar shear~$\chi^{(1)}$ as
\begin{equation}
{\kappa_1^{(1)}(t,\xvec)\over Hf_1} = \delta_1^{(1)}(t,\xvec)=
-{\Delta\II(\xvec)\over\HH^2f_1\Sigma_1}
\qquad \text{and} \qquad
\chi^{(1)}(t,\xvec) = -a^2\Delta\kappa^{(1)} = {\II(\xvec)\over H\Sigma_1}~.
\end{equation}

To compute the RHS of~\eqref{eq:main} to the second order in perturbations, first 
we need to derive the second-order curvature potential
by analytically integrating~\eqref{eq:time} over time:
\begin{equation}
\label{eq:pot2}
\CP^{(2)}(t,\xvec)= \II^{(2)}(\xvec) - \frac1{2\HH^2f_1\Sigma_1}\left[\frac12\nabla^i\II\nabla_i\II
+\Delta^{-1}\nabla^i\left(\nabla_i\II\Delta\II\right)\right]
\equiv \II^{(2)} + \varphi_2^{(2)} \, ,
\end{equation}
where $\II^{(2)}(\xvec)$ is the initial condition at the second order
and the quadratic terms are evaluated at the linear order in perturbations.
The curvature potential grows in time at the second order in proportion 
to the growth factor~$D_1(t)$, but they still vanish on superhorizon
scales.

Following the strategy in Section~\ref{ssec:plan}, the RHS of~\eqref{eq:main} 
at the second order in perturbations is written as
\begin{align}
\label{eq:2ndRHS}
\text{RHS}^{(2)} & = -\Delta\calR^{(2)} + \frac{3}{2}\nabla^i\calR\nabla_i\calR + 4\calR\Delta\calR + \frac{1}{\calH^2f_1\Sigma_1} \frac{\Delta}{2} \left[ \frac{1}{2}\nabla^i\calR\nabla_i\calR + \Delta^{-1}\nabla_i \left( \nabla^i\calR\Delta\calR \right) \right]
\nonumber\\
& \quad + \frac{1}{\calH^2\Sigma_1^2} \frac{1}{4} \left[ (\Delta\calR)^2 - \nabla^i\nabla^j\calR\nabla_i\nabla_j\calR \right] + \frac{1}{\calH^2f_1\Sigma_1^2} \left[ (\Delta\calR)^2 + \nabla^i\calR\Delta\nabla_i\calR \right] \, .
\end{align}
So there are four different time dependences, including 
$T_1=1$ for the first 3 terms on the RHS of \eqref{eq:2ndRHS} 
that leads to the linear-order growth factor $D_1(t)$ 
given by \eqref{eq:D1}. Thus, other than $D_1$, 
we find the new second-order growth factors for $\delta^{(2)}$ as
\begin{equation}
\label{eq:D2s}
D_{2A} = \frac{7}{5}H \int dt D_1^2f_1\Sigma_1 = \frac{D_1^2f_1\Sigma_1}{f_{2A}\Sigma_{2A}} \, ,
\quad
D_{2B} = \frac{7}{2}H \int dt D_1^2f_1^2 = \frac{D_1^2f_1^2}{f_{2B}\Sigma_{2B}} \, ,
\quad
D_{2C} = \frac{7}{2}H \int dt D_1^2f_1 = \frac{D_1^2f_1}{f_{2C}\Sigma_{2C}} \, ,
\end{equation}
with the corresponding time-independent spatial functions
\begin{equation}
\label{eq:2nd-D1}
X_1^{(2)} = -\Delta\calR^{(2)} + \frac{3}{2}\nabla^i\calR\nabla_i\calR + 4\calR\Delta\calR 
\end{equation}
for $D_1(t)$ and for $D_{2I}(t)$
\begin{align}
X_{2A}^{(2)} & = \frac{5}{14} \left[ \nabla_i \left( \nabla^i\calR\Delta\calR \right) + \frac{\Delta}{2} \left( \nabla^i\calR\nabla_i\calR \right) \right] \, ,
\\
X_{2B}^{(2)} & = \frac{1}{14} \left[ \nabla_i \left( \nabla^i\calR\Delta\calR \right) - \frac{\Delta}{2} \left( \nabla^i\calR\nabla_i\calR \right) \right] \, ,
\\
X_{2C}^{(2)} & = \frac{2}{7} \nabla_i \left( \nabla^i\calR\Delta\calR \right) \, .
\end{align}
Thus, the total second-order solution associated with RHS$^{(2)}$ is
\begin{equation}
\label{eq:sol-2nd}
\delta^{(2)}(t,\xvec) = \delta_1^{(2)} + \sum_{I=A}^C \delta_{2I}^{(2)} = D_1X_1^{(2)} + \sum_{I=A}^C D_{2I}X_{2I}^{(2)} \, .
\end{equation}
Note that not all $D_{2I}$'s are independent but they are subject to the constraint $D_{2A}+D_{2C} = 2D_1^2$. This allows us to rearrange $\delta_2^{(2)} \equiv \sum_I \delta_{2I}^{(2)}$ the same as the standard Newtonian form:
\begin{equation}
\delta_2^{(2)}(t,\xvec) = \frac{5D_{2A}+D_{2B}+4D_{2C}}{10} \left[ \frac{5}{7} \nabla_i \left( \nabla^i\calR\Delta\calR \right) \right] + \frac{5D_{2A}-D_{2B}}{4} \left[ \frac{\Delta}{7} \left( \nabla^i\calR\nabla_i\calR \right) \right] \, .
\end{equation}
Note that the two pure spatial functions inside the square brackets exactly correspond to the Newtonian second-order kernels $A_2(\kvec)$ and $B_2(\kvec)$ in the Fourier space: see \eqref{eq:Newtoniankernels}.

The second-order extrinsic curvature perturbation $\kappa^{(2)}$ and the scalar shear $\chi^{(2)}$ can be computed from the ADM energy constraint \eqref{eq:ADMene} and the momentum constraint equation~\eqref{eq:momentum} respectively, resulting
\begin{equation}
\begin{split}
\kappa^{(2)} & = \dot\delta_1^{(2)} + \frac{1}{2}\frac{d}{dt} \left\{ 2\sum_{I=A}^C \delta_{2I}^{(2)} - D_1^2 \left[ \nabla^i\calR\nabla_i\Delta\calR + (\Delta\calR)^2 \right] \right\} \equiv \dot\delta_1^{(2)} + \frac{\dot\calK}{2} \, ,
\\
\chi^{(2)} & = \frac{1}{H\Sigma_1} \left[ \calR^{(2)} - \calR^2 - \frac{1}{2}\nabla^i\calR\nabla_i\calR + \frac{3}{2}\Delta^{-2} \nabla_i\nabla_j \left( \nabla^i\calR\nabla^j\calR \right) \right] - a^2\Delta^{-1} \kappa_2^{(2)} \equiv \chi_1^{(2)}  - \frac{a^2}{2} \Delta^{-1} \dot\calK \, .
\end{split}
\end{equation}
Note from above that $H\Sigma_1\chi_1^{(2)}$ constant, and that $\calK$ contains four different time dependences: $D_{2A}$, $D_{2B}$, $D_{2C}$ and $D_1^2$, which all become identical to $D_1^2$ in the EdS universe.

\subsection{Third-order solutions}
\label{ssec:third}

At the third order in perturbations, we need to consider cubic terms, 
consisting of three perturbation variables evaluated at the linear order 
to make the cubic term at the third order. In addition, we need to consider
quadratic terms that were evaluated in the previous section at the second 
order,
because those quadratic terms also contribute to the third order with
one variable at the second order and the other at the linear order.

Following the same strategy in Section~\ref{ssec:known-sol}, we first integrate
\eqref{eq:time} to derive the third-order curvature potential. We then find
\begin{align}\label{eq:varphi3}
\varphi^{(3)}(t,\xvec) & = \calR^{(3)}(\xvec) + 2\calR\calR^{(2)} + 2\calR\varphi_2^{(2)} + 2D_1 \Delta^{-1} \nabla_i \left( \nabla^i\calR\nabla^j\calR\nabla_j\calR + \calR\nabla^i\nabla^j\calR\nabla_j\calR + \calR\Delta\calR\nabla^i\calR \right) 
\nonumber\\
& \quad - \frac{D_1}{2}\Delta^{-1} \nabla_i \left( \nabla^i\nabla^j\calR^{(2)}\nabla_j\calR + \Delta\calR^{(2)}\nabla^i\calR \right) - \frac{D_1}{4}\Delta^{-1} \nabla_i \left( \nabla^i\nabla^j\varphi_2^{(2)}\nabla_j\calR + \Delta\varphi_2^{(2)}\nabla^i\calR \right)
\nonumber\\
& \quad - \frac{D_1}{2}\Delta^{-1}\nabla_i \left[ \nabla^i\nabla^j\calR \nabla_j \left( H\Sigma_1\chi_1^{(2)} \right) + \Delta\calR \nabla^i \left( H\Sigma_1\chi_1^{(2)} \right) \right] + \frac{\Delta^{-1}}{4} \nabla_i \left( \nabla^i\nabla^j\calR\Delta^{-1}\nabla_j\calK + \Delta\calR\Delta^{-1}\nabla^i\calK \right) \, ,
\end{align}
where $\calR^{(3)}(\xvec)$ is a pure third order integration constant. With the third-order curvature potential, the RHS of~\eqref{eq:main} 
at the third order in perturbations is then
\begin{align}
\label{eq:rhs3rd}
\up{RHS}^{(3)}(t,\xvec) & = -\Delta\varphi^{(3)} + 3\nabla^i\varphi_2^{(2)}\nabla_i\calR + 4\varphi_2^{(2)}\Delta\calR + 4\calR\Delta\varphi_2^{(2)}
\nonumber\\
& \quad + 3\nabla^i\calR^{(2)}\nabla_i\calR + 4\calR^{(2)}\Delta\calR + 4\calR\Delta\calR^{(2)} - 3\calR \left( 3\nabla^i\calR\nabla_i\calR + 4\calR\Delta\calR \right) 
\nonumber\\
& \quad - a^2H \left[ -D_1\Delta \nabla_i \calR \nabla^i \left( \frac{\chi_1^{(2)}}{a^2} - \frac{\Delta^{-1}}{2}\dot\calK \right) + \dot{D}_1\nabla_i\calR \nabla^i\delta^{(2)} \right] - 2a^2HD_1\dot{D}_1\calR\nabla^i\calR\Delta\nabla_i\calR
\nonumber\\
& \quad - a^2H \Delta\calR \left[ \dot{D}_1\delta^{(2)} + D_1 \left( \dot\delta_1^{(2)} + \frac{\dot\calK}{2} \right) \right]
+ \frac{a^2}{2}\dot{D}_1 \left[ \frac{\Delta\calR}{3} \left( \frac{\Delta\chi_1^{(2)}}{a^2} - \frac{\dot\calK}{2} \right) - \nabla_i\nabla_j\calR \nabla^i\nabla^j \left( \frac{\chi_1^{(2)}}{a^2} - \frac{\Delta^{-1}}{2}\dot\calK \right) \right]
\nonumber\\
& \quad + a^2\dot{D}_1^2 \left[ \calR\nabla^i\nabla^j\calR\nabla_i\nabla_j\calR + \nabla^i\calR\nabla^j\calR\nabla_i\nabla_j\calR - \frac{1}{3}\calR(\Delta\calR)^2 - \frac{1}{3}\nabla^i\calR\nabla_i\calR\Delta\calR \right]
- \frac{\Delta\calR}{3H\Sigma_1} \left( \dot\delta_1^{(2)} + \frac{\dot\calK}{2} \right) \, .
\end{align}

From these we can first find the time-independent spatial function on the RHS which give rise to $D_1$:
\begin{equation}
\label{eq:3rd-D1}
X_1^{(3)} = -\Delta\calR^{(3)} -2\Delta\left( \calR\calR^{(2)} \right) + 3\nabla^i\calR^{(2)}\nabla_i\calR + 4\calR^{(2)}\Delta\calR + 4\calR\Delta\calR^{(2)} - 3\calR \left( 3\nabla^i\calR\nabla_i\calR + 4\calR\Delta\calR \right) \, ,
\end{equation}
which leads to the third-order extrinsic curvature perturbation proportional to $D_1$:
\begin{equation}
\label{eq:kkk1}
\frac{\kappa_1^{(3)}(t,\xvec)}{Hf_1} = \delta_1^{(3)}(t,\xvec) = D_1(t)X_1^{(3)}(\xvec) \, .
\end{equation}
Those associated with the growth factors $D_{2A}$, $D_{2B}$ and $D_{2C}$ found in Section~\ref{ssec:known-sol}:
\begin{align}
X_{2A}^{(3)} & = \frac{5}{7} \left\{ -\frac{\nabla^i\varphi_2^{(2)}}{D_1}\nabla_i\calR + 2\frac{\varphi_2^{(2)}}{D_1}\Delta\calR + 2\calR\frac{\Delta\varphi_2^{(2)}}{D_1} + \nabla_i \bigg[ -2\nabla^i\calR\nabla^j\calR\nabla_j\calR - 2\calR\nabla^i\nabla^j\calR\nabla_j\calR - 2\calR\Delta\calR\nabla^i\calR  \right.
\nonumber\\
& \hspace{10em} \left.\left. + \frac{1}{2}\nabla^i\nabla^j\calR^{(2)}\nabla_j\calR + \frac{1}{2}\Delta\calR_2\nabla^i\calR + \frac{1}{2}\nabla^i\nabla^j\calR\nabla_j \left( H\Sigma_1\chi_1^{(2)} \right) + \frac{1}{2}\Delta\calR \nabla^i \left( H\Sigma_1\chi_1^{(2)} \right) \right] \right\} \, ,
\\
X_{2B}^{(3)} & = \frac{2}{7} \left[ \frac{\Delta\calR}{6}\Delta \left( H\Sigma_1\chi_1^{(2)} \right) - \frac{\nabla_i\nabla_j\calR}{2} \nabla^i\nabla^j \left( H\Sigma_1\chi_1^{(2)} \right) \right.
\nonumber\\
& \qquad\quad \left. + \calR\nabla^i\nabla^j\calR\nabla_i\nabla_j\calR + \nabla^i\calR\nabla^j\calR\nabla_i\nabla_j\calR - \frac{1}{3}\calR(\Delta\calR)^2 - \frac{1}{3}\nabla^i\calR\nabla_i\calR\Delta\calR - \frac{\Delta\calR}{3}X_1^{(2)} \right] \, ,
\\
X_{2C}^{(3)} & = \frac{2}{7} \left[ \Delta\nabla_i \calR \nabla^i \left( H\Sigma_i\chi_1^{(2)} \right) - 2\calR\nabla^i\calR\Delta\nabla_i\calR - \nabla_i\calR\nabla^iX_1^{(2)} - 2\Delta\calR X_1^{(2)} \right] \, .
\end{align}
Then we find the third-order solution with the second-order growth factors as
\begin{equation}
\label{eq:sol-3rd_2}
\delta_2^{(3)}(t,\mathbi{x}) = \sum_{I=A}^C D_{2I}(t)X_2I^{(3)}(\xvec) \, .
\end{equation}
We also have new growth factors that become $D_1^3$ in the EdS universe:
\beeq
D_{3D}  \equiv \frac{9}{5}H \int dt D_1^3f_1\Sigma_1 \, ,
\qquad\qquad
D_{3E}  \equiv \frac{9}{2}H \int dt D_1^3f_1 \, ,
\qquad\qquad
D_{3F}  \equiv \frac{9}{2}H \int dt D_1^3f_1^2 \, ,
\eneq
with the spatial functions associated with them:
\begin{align}
X_{3D}^{(3)} & = \frac{5}{36} \nabla_i \left( \frac{\nabla^i\nabla^j\varphi_2^{(2)}}{D_1}\nabla_j\calR + \frac{\Delta\varphi_2^{(2)}}{D_1}\nabla^i\calR \right) + \frac{5}{36} \nabla_i \left\{ \left( \nabla^i\nabla^j\calR \Delta^{-1}\nabla_j + \Delta\calR \Delta^{-1}\nabla^i \right) \left[ \nabla^k\calR\Delta\nabla_k\calR + (\Delta\calR)^2 \right] \right\} \, ,
\\
X_{3E}^{(3)} & = \frac{1}{9} \left\{ \Delta^{-1} \nabla^i \left[ \nabla^j\calR\Delta\nabla_j\calR + (\Delta\calR)^2 \right] \Delta\nabla_i\calR + \left[ \nabla^j\calR\Delta\nabla_j\calR + (\Delta\calR)^2 \right] \Delta\calR \right\} \, ,
\\
X_{3F}^{(3)} & = -\frac{1}{18} \left( \nabla^i\nabla^j\calR \Delta^{-1}\nabla_i\nabla_j - \Delta\calR \right) \left[ \nabla^k\calR\Delta\nabla_k\calR + (\Delta\calR)^2 \right] \, .
\end{align}
The associated third-order solution, which constitutes one part of $\delta_3^{(3)}$ is
\begin{equation}
\label{eq:sol-3rd_3-1}
\delta_3^{(3)}(t,\xvec) \supset \sum_{I=D}^F D_{3I}(t)X_{3I}^{(3)}(\xvec) \, .
\end{equation}
Finally, the growth factors coming from $\delta_{2I}^{(2)}$ ($I=A,B,C$) also scale as $D_1^3$ in the EdS universe:
\begin{align}
D_{3Ia} & \equiv \frac{9}{5} H \int dt D_1f_1\Sigma_1D_{2I} \, ,
\qquad\qquad
D_{3Ib} \equiv \frac{9}{4} H \int dt D_1D_{2I}f_{2I} \, ,
\\
D_{3Ic} & \equiv \frac{9}{2} H \int dt D_1f_1D_{2I} \, ,
\qquad\qquad
D_{3Id}  \equiv \frac{9}{4} H \int dt D_1f_1D_{2I}f_{2I} \, ,
\end{align}
with the corresponding spatial functions:
\begin{align}
X_{3Ia}^{(3)} & = -\frac{5}{18} \nabla_i \left[ \left( \nabla^i\nabla^j\calR \Delta^{-1}\nabla_j + \Delta\calR \Delta^{-1}\nabla^i \right) X_{2I}^{(2)} \right] \, ,
\\
X_{3Ib}^{(3)} & = -\frac{4}{9} \nabla_i \left( \Delta\calR \Delta^{-1}\nabla^iX_{2I}^{(2)} \right) \, ,
\\
X_{3Ic}^{(3)} & = -\frac{2}{9} \nabla_i \left( X_{2I}^{(2)} \nabla^i\calR \right) \, ,
\\
X_{3Id}^{(3)} & = \frac{2}{9} \left( \nabla^i\nabla^j\calR \Delta^{-1}\nabla_i\nabla_j - \Delta\calR \right) X_{2I}^{(2)} \, .
\end{align}
These give the other part of $\delta_3^{(3)}$:
\begin{equation}
\label{eq:sol-3rd_3-2}
\delta_3^{(3)}(t,\xvec) \supset \sum_{I=A}^C \sum_{i=a}^d D_{3Ii}(t)X_{3Ii}^{(3)}(\xvec) \, .
\end{equation}
The full third-order solution is the sum of \eqref{eq:kkk1}, \eqref{eq:sol-3rd_2}, \eqref{eq:sol-3rd_3-1} and \eqref{eq:sol-3rd_3-2}:
\begin{equation}
\label{eq:sol-3rd}
\delta^{(3)}(t,\xvec) = \delta_1^{(3)} + \delta_2^{(3)} + \delta_3^{(3)}
= D_1X_1^{(3)} + \sum_{I=A}^C D_{2I}X_{2I}^{(3)} + \sum_{I=D}^F D_{3I}X_{3I}^{(3)} + \sum_{I=A}^C \sum_{i=a}^d D_{3Ii}X_{3Ii}^{(3)} \, .
\end{equation}
This analytic third-order solution is one of the main results of this article.

\subsection{Full third-order solutions in the EdS universe}
\label{ssec:eds}

In the EdS universe, it is only the matter density that
drives the Hubble expansion and the growth of perturbations, thus providing
the simplest example and consistency checks, 
to which we can compare our analytic solutions in a $\Lambda$CDM universe.

With $\Omega_m=1$, the Hubble parameter is $H=2/(3t)$, and all the quantities are
scale-free in their time-dependence.
The RHS of~\eqref{eq:main} has the simple time-dependence:
\beeq
\up{RHS}^{(n)}(t,\xvec)\propto{1\over\HH^{2(n-1)}}~;
\qquad
T_1(t)=1~,
\qquad 
T_2(t)={14\over25\HH^2}~,
\qquad 
T_3(t)={36\over75\HH^4}~,
\eneq
regardless of its order in perturbations. Therefore, all the growth factors
associated with each RHS$^{(n)}$ are all identical, and they
can be analytically integrated as
\begin{alignat}{3}
D_1(t)&=H\int dt~{1\over\HH^2}={2\over5\HH^2}~,
\qquad 
& f_1 =1~,
\qquad
\Sigma_1 &=\frac52~,
\\
D_2(t)&=H\int dt~{14\over25\HH^4}={2^2\over5^2\HH^4}=D_1^2~,
\qquad 
& f_2 =2~,
\qquad
\Sigma_2 & =\frac74~,
\\
D_3(t)&=H\int dt~{36\over75\HH^6}={2^3\over5^3\HH^6}=D_1^3~,
\qquad 
& f_3 =3~,
\qquad
\Sigma_3 & =\frac32~.
\end{alignat}
The Newtonian
solutions in the EdS universe are then
\begin{align}
\label{eq:newtsol}
\delta_1^{(1)}(t,\xvec)&={\kappa_1^{(1)}(t,\xvec)\over H}
=-D_1(t)\Delta\II(\xvec)~,
\\
\label{eq:newtsol2}
\delta_2^{(2)}(t,\xvec)&={D_1^2(t)\over7}\bigg[5(\Delta\II)^2
+2\nabla^i\nabla^j\II\nabla_i\nabla_j\II+7\nabla_i\II\Delta\nabla^i\II\bigg]~,
\\
\label{eq:newtsol2k}
{\kappa_2^{(2)}(t,\xvec)\over H}&={D_1^2(t)\over7}\bigg[3(\Delta\II)^2
+4\nabla^i\nabla^j\II\nabla_i\nabla_j\II+7\nabla_i\II\Delta\nabla^i\II\bigg]~,
\\
\delta_3^{(3)}(t,\xvec)&=
-{D_1(t)\over18}\left[
2\Delta\left(\nabla_i\II\Delta^{-1}\nabla^i{\kappa_2^{(2)}\over H}\right)
+7\nabla^i\left(\Delta^{-1}\nabla_i{\kappa^{(2)}_2\over H}\Delta\II\right)
+7\nabla^i\left(\delta_2^{(2)}\nabla_i\II\right)\right]~,
\\
\label{eq:newtsol3}
{\kappa_3^{(3)}(t,\xvec)\over H}&=
-{D_1(t)\over6}\left[
2\Delta\left(\nabla_i\II\Delta^{-1}\nabla^i{\kappa_2^{(2)}\over H}\right)
+\nabla^i\left(\Delta^{-1}\nabla_i{\kappa^{(2)}_2\over H}\Delta\II\right)
+\nabla^i\left(\delta_2^{(2)}\nabla_i\II\right)\right]~,
\end{align}
and the relativistic solutions are
\begin{align}
\label{eq:relsol}
\delta_1^{(2,3)}(t,\xvec)&={\kappa_1^{(2,3)}(t,\xvec)\over H}=D_1(t)
\left[\frac32\nabla^i\II\nabla_i\II+4\II\Delta\II
-3\II\left(3\nabla^i\II\nabla_i\II+4\II\Delta\II\right)\right]~,
\\
\label{eq:relsol3d}
\delta_2^{(3)}(t,\xvec)&={D_1^2(t)\over7}
\Bigg[ \frac83\II(\Delta\II)^2
-8\II\nabla^i\nabla^j\II\nabla_i\nabla_j\II-14\II
\nabla^i\II\Delta\nabla_i\II
-8\nabla^i\nabla^j\II\nabla_i\II\nabla_j\II 
\nnn
& \qquad\qquad
+{8\over3}\nabla^i\II\nabla_i\II\Delta\II
+\left(7\Delta\nabla_i\II\nabla^i+4\nabla_i\nabla_j\II\nabla^i\nabla^j
-{4\over3}\Delta\II\Delta\right)\left( 
D_1^{-1}\Delta^{-1}\delta_1^{(2)} + \frac{5}{2}H\Delta\chi_1^{(2)}
\right) 
\nnn
& \qquad\qquad
-\left(10\Delta\II+7\Delta\nabla_i\II\Delta^{-1}\nabla^{i}
+7\nabla_i\II\nabla^{i}+4\nabla_i\nabla_j\II\Delta^{-1}\nabla^i\nabla^j\right)
D_1^{-1}\delta_1^{(2)}\Bigg]~,
\\
\label{eq:relsol3}
{\kappa_2^{(3)}(t,\xvec)\over H}&={D_1^2(t)\over7}
\Bigg[\frac{16}3\II(\Delta\II)^2
-16\nabla^i\nabla^j\II\nabla_i\nabla_j\II
-14\nabla^i\II\Delta\nabla_i\II
-16\nabla^i\nabla^j\II\nabla_i\II\nabla_j \II
\nnn
& \qquad\qquad
+{16\over3}\nabla^i\II\nabla_i \II\Delta\II
+\left(7\Delta\nabla_i\II\nabla^i+8\nabla_i\nabla_j\II\nabla^i\nabla^j
-{8\over3}\Delta\II\Delta\right)\left( 
D_1^{-1}\Delta^{-1}\delta_1^{(2)} + \frac{5}{2}H\Delta\chi_1^{(2)}\right)
\nnn
& \qquad\qquad
-\left(6\Delta\II+7\Delta\nabla_i\II\Delta^{-1}\nabla^i+7\nabla_i\II\nabla^i
+8\nabla_i\nabla_j\II\Delta^{-1}\nabla^i\nabla^j\right)
D_1^{-1}\delta_1^{(2)}\Bigg]~.
\end{align}
In presenting the above solutions, we assumed the initial 
condition~$\II(\xvec)$ is at the linear order in perturbations. But in 
principle, the initial condition can be treated as a nonlinear perturbation
variable, such that $\delta_1=-D_1\Delta\II$ also contributes
to $n$-th order in perturbation, for instance, if $\II=\II^{(n)}$
as is explicit in \eqref{eq:2nd-D1} and \eqref{eq:3rd-D1}.

\section{Comparison to previous works in general relativity}
\label{sec:comp}
\setcounter{equation}{0}

In this section, we compare our analytic solutions in Section~\ref{sec:solution}
to the solution derived in \cite{NOHW08,JEGOET11,Biern:2014zja}.
The full relativistic matter density fluctuation and its one-loop power
spectrum \cite{NOHW08,JEGOET11} as well as one-loop bispectrum 
\cite{Biern:2014zja} were computed under the same gauge condition,
but by assuming the EdS universe. However, the solution in
\cite{NOHW08,JEGOET11,Biern:2014zja} differs from ours in the relativistic corrections.
The critical differences in the
previous works are that (a) the initial condition is set by the density 
fluctuation rather than the curvature perturbation~$\CP$,
and that (b) it was assumed to be at the linear order in perturbation, i.e.,
$\delta(\xvec,t_i)=\delta_1^{(1)}(\xvec,t_i)$ at some early time~$t_i$.
Since the curvature perturbation
spectrum is set up by inflation in the early Universe and
is conserved on super-horizon scales, it is more natural to set up the initial
condition for the nonlinear evolution of the matter density fluctuation
with the curvature perturbation as in the current study. However, their set-up
with~$\delta$ is just fine, as we show below that the ADM energy constraint 
equation~(\ref{eq:conv}) 
leads to the equivalent initial condition set up by the curvature perturbation.
The real difference lies in (b),
in clear disagreement with our
finding in Section~\ref{sec:solution}. Here we derive the previous work in
configuration space, rather than in Fourier space as was done in 
\cite{NOHW08,JEGOET11,Biern:2014zja}, and show how these differences play a role in
connecting two solutions.

\subsection{Derivation of previous works in configuration space}
\label{ssec:EdSmain}

While the relativistic dynamical equations are identical, the approach to the solution in previous works focuses on the main dynamical
variables~$\delta$ and~$\kappa$, rather than~$\II$, which play a role
of supplementing the Newtonian dynamical equations with relativistic corrections.
The conservation equation~\eqref{eq:mast1} and the Raychaudhuri 
equation~\eqref{eq:mast2} are explicitly expanded
up to the third order in perturbations by using 
\eqref{eq:ADMNi} and~\eqref{eq:shear} as
\begin{align}
\label{eq:dyn1}
\dot\delta-\kappa&=-{1\over a^2}(1-2\CP)\nabla^i\chi\nabla_i\delta
+\delta\kappa~,
\\
\label{eq:dyn2}
\dot\kappa+2H\kappa-4\pi G\bar\rho_m\delta
&=-{1\over a^2}(1-2\CP)\nabla^i\chi\nabla_i\kappa+\frac13\kappa^2
+{1 \over a^4} \left[\nabla_i\nabla_j\chi
\nabla^i\nabla^j \chi-{1 \over 3}\left(\Delta \chi\right)^2 \right](1-4\CP)
\nnn
& \quad
-{4\over a^4}\left( \nabla^i\nabla^j\chi
- {1 \over 3}\gbar^{ij} \Delta \chi \right)\nabla_i\chi\nabla_j\CP~.
\end{align}
These dynamical equations are solved in conjunction with the ADM energy 
constraint equation at the linear order
\beeq
\label{eq:admlin}
\frac32H^2\delta+H\kappa+{\Delta\over a^2}\CP=0~,
\eneq
and the second-order ADM momentum constraint equation~(\ref{eq:momentum}). 
We will only use the linear-order curvature perturbation~$\CP$
and the second-order scalar shear~$\chi$ in this section, complementing
the dynamical equations for~$\delta$ and~$\kappa$.

With the knowledge of the time dependence of the EdS solutions
in Section~\ref{ssec:eds}, we seek
solutions of the dynamical equations~(\ref{eq:dyn1}) and~(\ref{eq:dyn2})
by explicitly removing their time-dependence.
Up to third order, the density fluctuation $\delta$ and the perturbation to the extrinsic curvature $\kappa$ are
parametrized as
\beeq
\delta(t,\xvec)\equiv{\SA_1(\xvec)\over\HH^2}+{\SA_2(\xvec)\over\HH^4}
+{\SA_3(\xvec)\over\HH^6}~,
\qquad\qquad
{\kappa(t,\xvec)\over H}\equiv{\SB_1(\xvec)\over\HH^2}+{\SB_2(\xvec)\over\HH^4}
+{\SB_3(\xvec)\over\HH^6}~,
\eneq
where $\SA_i(\xvec)$ and $\SB_i(\xvec)$ are time-independent spatial
functions and they vanish when the $n$-th order perturbation is considered,
if $i>n$. 
Note that we have grouped $\delta$ and $\kappa$ according to the same 
time dependence, {\em not} to the perturbation order, as denoted by subscripts.
For example, $c_1(\xvec)$ may contain non-linear perturbation terms 
as we will show right below.
Inspecting the ADM momentum constraint  equation~(\ref{eq:momentum}),
we can also parametrized the scalar shear up to the second order in 
perturbations as
\beeq
H\chi\equiv\SC_1(\xvec)+{\SC_2(\xvec)\over\HH^2}~,
\eneq
and of course the curvature perturbation~$\CP(\xvec)$ 
is time-independent at the  linear order.
Comparing to the notation in Section~\ref{sec:solution}, we can readily identify
the correspondence:
\beeq
\delta_i(t,\xvec)={c_i(\xvec)\over\HH^{2i}}~,\qquad \qquad 
\chi_1={e_1(\xvec)\over H}~,\qquad \qquad
\chi_2={e_2(\xvec)\over a^2H^3}~.
\eneq

Armed with the parametrized solutions, the conservation 
equation~(\ref{eq:dyn1}) yields a set of {\em algebraic} equations
\begin{equation}
\begin{split}
\SA_1-\SB_1&=0~,
\\
2\SA_2-\SB_2&=-(1-2\CP)\nabla^i\SC_1\nabla_i\SA_1+\SA_1\SB_1~,
\\
3\SA_3-\SB_3&=-(1-2\CP)\left(\nabla^i\SC_1\nabla_i\SA_2+
\nabla^i\SC_2\nabla_i\SA_1\right)+\SA_1\SB_2+\SA_2\SB_1~,
\end{split}
\end{equation}
and we now simply set $\SA_1=\SB_1\equiv\SD$ at all orders in perturbation.
Similarly, the Raychaudhuri equation~(\ref{eq:dyn2}) provides
\begin{equation}
\begin{split}
{5\SB_2-3\SA_2\over2}&=
-(1-2\CP)\nabla^i\SC_1\nabla_i\SB_1
+\frac13\SB_1^2+(1-4\CP)\left[\nabla_i\nabla_j\SC_1\nabla^i\nabla^j\SC_1
-\frac13(\Delta\SC_1)^2\right] 
\\
& \quad
-4\left( \nabla^i\nabla^j\SC_1
- {1 \over 3}\gbar^{ij} \Delta \SC_1 \right)\nabla_i\SC_1\nabla_j\CP~,
\\
{7\SB_3-3\SA_3\over2}&=
-\nabla^i\SC_1\nabla_i\SB_2-
\nabla^i\SC_2\nabla_i\SB_1+\frac23\SB_1\SB_2
+2\nabla^i\nabla^j\SC_1
\nabla_i\nabla_j\SC_2-\frac23\Delta\SC_1\Delta\SC_2~,
\end{split}
\end{equation}
and the ADM momentum constraint provides the supplementary equation for
the scalar shear\footnote{
Here, $\Psi$ is identical to the following second-order quantity:
\begin{equation*}
\Psi^{(2)}(\xvec) \equiv \frac{\delta_1^{(2)}(t,\xvec)}{D_1(t)} + H\Sigma_1\Delta\chi_1^{(2)}(t,\xvec) \, .
\end{equation*}
}
\begin{equation}
\label{eq:scalarc}
\begin{split}
\SB_1+\Delta\SC_1&=
2\CP\Delta\SC_1-\nabla^i\CP\nabla_i\SC_1+{3\over2}\Delta^{-1}\nabla^i
\left(\nabla_i\SC_1\Delta\CP+\nabla_j\nabla_i\CP\nabla^j\SC_1\right)
\equiv\Delta\Psi~,
\\
\SB_2+\Delta\SC_2&=0~.
\end{split}
\end{equation}
Therefore, the solutions to the algebraic equations are
\begin{align}
\label{eq:conf1}
\SA_2 &=\frac17\SD^2\left(5+\frac83\CP\right)
+(1-2\CP)\nabla^i\Delta^{-1}\SD\nabla_i\SD
+\frac27(1-4\CP)\nabla_i\nabla_j\Delta^{-1}\SD\nabla^i\nabla^j\Delta^{-1}\SD
\nnn
& \quad
-\frac87\left( \nabla^i\nabla^j\Delta^{-1}\SD- {1 \over 3}\gbar^{ij}\SD\right)
\nabla_i\Delta^{-1}\SD\nabla_j\CP
-\left(\nabla_i\SD\nabla^i+\frac47\nabla_i\nabla_j\Delta^{-1}\SD
\nabla^i\nabla^j-{4\over21}\SD\Delta\right)\Psi~,
\\
\SA_3
&={1\over18}\bigg[7\nabla^i\left(\SA_2\Delta^{-1}\nabla_i\SD\right)
+2\Delta\left(\Delta^{-1}\nabla_i\SD\nabla^i\Delta^{-1}
\SB_2\right)+7\nabla^i\left(\SD\nabla_i\Delta^{-1}\SB_2\right)\bigg]~,
\\
\SB_2 &=\frac17\SD^2\left(3+{16\over3}\CP\right)
+(1-2\CP)\nabla^i\Delta^{-1}\SD\nabla_i\SD
+\frac47(1-4\CP)\nabla_i\nabla_j\Delta^{-1}\SD\nabla^i\nabla^j\Delta^{-1}\SD
\nnn
& \quad
-{16\over7}\left( \nabla^i\nabla^j\Delta^{-1}\SD- {1 \over 3}\gbar^{ij}\SD
\right)\nabla_i\Delta^{-1}\SD\nabla_j\CP
-\left(\nabla_i\SD\nabla^i+\frac87\nabla_i\nabla_j\Delta^{-1}\SD\nabla^i
\nabla^j-{8\over21}\SD\Delta\right)\Psi~,
\\
\label{eq:conf2}
\SB_3
&=\frac16\bigg[\nabla^i\left(\SA_2\Delta^{-1}\nabla_i\SD\right)
+2\Delta\left(\Delta^{-1}\nabla_i\SD\nabla^i\Delta^{-1}
\SB_2\right)+\nabla^i\left(\SD\nabla_i\Delta^{-1}\SB_2\right)\bigg]~.
\end{align}
This completes our derivation of previous work in configuration space.
It is noted that the solution was derived \cite{NOHW08,JEGOET11,Biern:2014zja} in Fourier
space (see Section~\ref{ssec:grkernel}), and the initial condition set up with
the coefficient~$\SD$ of the density fluctuation is assumed to be at the
linear-order in perturbations.

\subsection{Comparison to our solution}
\label{ssec:compour}

Compared to our analytic solutions in 
\eqref{eq:newtsol}$-$\eqref{eq:relsol3}, these solutions are
expressed in terms of the coefficient~$\SD(\xvec)$ of the density fluctuation
in proportion to~$1/\HH^2$, rather than the curvature perturbation~$\II$.
Given the ADM energy constraint equation~(\ref{eq:admlin}) at the {\em linear
order} in perturbation, the relation of this coefficient to the curvature
perturbation is
\beeq
\SD=\SA_1=\SB_1=-{2\over5}\Delta\CP=-\frac25\Delta\II~.
\eneq
With this relation, we can easily 
recover our analytic solutions~$\delta_i^{(i)}$
and~$\kappa_i^{(i)}$ in~\eqref{eq:newtsol}$-$\eqref{eq:newtsol3}, 
corresponding to the standard
Newtonian solutions, while there remain the differences in the relativistic
corrections~$\delta_i^{(j)}$ and~$\kappa_i^{(j)}$ for $j>i$.

The reason for this difference is that the density fluctuation~$\delta_1$
(or the coefficient~$\SD$) is treated as the linear-order perturbation $\delta^{(1)}$ ---
in our derivation we made {\em no assumption} about the perturbation orders
of all the coefficients, while only
the linear-order ADM energy constraint is used to convert the parametrized
solutions and compare to our analytic solutions in Section~\ref{sec:solution}.
Using the full ADM energy constraint equation~(\ref{eq:enecons}) and 
taking the limit $t\RA0$, we derive the nonlinear relation
\beeq
\label{eq:conv}
\SD(\xvec)
=\frac25\left[-\Delta\II+{3\over2}\nabla^i\II\nabla_i\II+4\II\Delta\II
-3\II(3\nabla^i\II\nabla_i\II+4\II\Delta\II)\right]~,
\eneq
and the solutions up to third order in perturbation in previous work are readily obtained as
\beeq
\delta_1={\kappa_1\over H}={\SD\over\HH^2}
={2\over5\HH^2}\left[-\Delta\II+{3\over2}\nabla^i\II\nabla_i\II
+4\II\Delta\II-3\II(3\nabla^i\II\nabla_i\II+4\II\Delta\II)\right]~,
\eneq
identical to our analytic solution in Section~\ref{sec:solution}.
Using the relation to the second order in perturbations, we can repeat
this exercise to recover the relativistic corrections 
in~$\delta_2^{(3)}$ and~$\kappa_2^{(3)}$. This proves the equivalence of
the solutions in~\eqref{eq:conf1}$-$\eqref{eq:conf2} 
to the analytic solutions in 
Section~\ref{sec:solution}. However, the density fluctuation~$\delta_1$ (or
the coefficient~$\SD$) is {\it not} a linear-order perturbation, as
is apparent in~(\ref{eq:conv}). Therefore, the Fourier kernels
derived in \cite{NOHW08,JEGOET11,Biern:2014zja} are valid {\em only under
 the assumption} that the initial condition set up by the density
fluctuation~$\delta_1$ is linear order in perturbations. We will provide
the complete Fourier kernels in Section~\ref{ssec:grkernel}.

\section{Comparison to the standard Newtonian perturbation theory}
\label{sec:newt}
\setcounter{equation}{0}

In this section, we compare our relativistic solutions to the standard
Newtonian solutions. This provides insights to understand the connection of 
the relativistic dynamics to the Newtonian dynamics. 
We also derive the Fourier kernels for the relativistic solutions.

\subsection{Dynamical equations of motion}
\label{ssec:dyn}

As noted in the previous sections, the relativistic solutions in our gauge
condition closely resemble the standard Newtonian ones. In the comoving
gauge, the dynamical equations of motion are shown to be identical to the 
Newtonian ones to the second order in perturbations \cite{HWNO99,HWNO05}
with the relativistic effects appearing only from the third order \cite{HWNO05b}.
This gauge condition is later shown \cite{YOO14b} to correspond to the
proper-time hypersurface of nonrelativistic matter flows. In the proper-time
hypersurface, the local observer moving with the nonrelativistic matter flows
can measure the energy density in its rest frame, providing the most natural
description of the matter density fluctuation.

Following this approach, we compare the relativistic dynamical
equations~(\ref{eq:dyn1}) and~(\ref{eq:dyn2}) with those in the Newtonian
dynamics by identifying proper correspondences between the relativistic
and Newtonian dynamics. In the standard Newtonian perturbation theory,
the velocity~$\vvec_N$ of the flow is often expressed in terms of the velocity divergence field $\theta_N$:
\beeq
\theta_N\equiv{1\over a}\nabla\cdot\vvec_N~,
\eneq
where the subscript~$N$ is used to indicate the quantity 
is a Newtonian variable.
Since in general relativistic approach with our gauge condition, $\kappa$ 
is the perturbation
in the expansion of the local observer, we define
the (nonlinear) relativistic velocity~$\vvec$ of the observer as
\beeq
-\kappa\equiv{1\over a}\nabla\cdot\vvec~.
\eneq
This ``velocity'' is defined only in relation to~$\kappa$, which is {\it not}
identical to the spatial component of the four velocity in 
\eqref{eq:four}.

With this identification of velocity, 
the relativistic dynamical equations~\eqref{eq:dyn1} and~\eqref{eq:dyn2}
can be rewritten in terms of the matter density fluctuation and the velocity
as
\begin{align}
\label{eq:Newt1}
& \dot\delta+ {1\over a}\nabla\cdot\vvec =
-{1\over a}\nabla\cdot\left(\vvec\delta\right)
+{2\CP\over a}\nabla\delta\cdot\vvec  
\nonumber\\
&\qquad\qquad\qquad
-{1\over a}\nabla\delta\cdot\nabla\Delta^{-1}
\left[2\CP\nabla\cdot\vvec-(\vvec\cdot\nabla)\CP
+{3\over2}\Delta^{-1}\nabla\cdot\Big((\vvec\cdot\nabla)
\nabla\CP+\vvec\Delta\CP\Big)\right]~, 
\\
\label{eq:Newt2}
& \nabla\cdot\dot\vvec+H\nabla\cdot\vvec+{3H^2\over2}a\Omega_m\delta =
-{1\over a}\nabla\cdot\left[\left(\vvec\cdot\nabla\right)\vvec\right]
-{2\over3a}\CP(\vvec\cdot\nabla)(\nabla \cdot\vvec)
+{4\over a}\nabla\cdot\left[\CP\left((\vvec\cdot\nabla)\vvec-{1 \over 3}
\vvec(\nabla\cdot\vvec)\right) \right] 
\nnn
& \qquad\qquad\qquad
+{1\over a}\left[\vvec\cdot\nabla+{2\over 3} \nabla \cdot \vvec
-\Delta\Big((\vvec\cdot\nabla)\Delta^{-1}\Big)\right]
\left[2\CP\nabla\cdot\vvec-(\vvec\cdot\nabla)\CP
+{3\over2}\Delta^{-1}\nabla\cdot\Big((\vvec\cdot\nabla)
\nabla\CP+\vvec\Delta\CP\Big)\right]~.
\end{align}
It is now evident that the relativistic dynamics in our gauge condition
with the proper correspondence between $(\delta,\vvec)$ and $(\delta_N,\vvec_N)$ 
follows the standard Newtonian equations of 
motion up to the second order in perturbations and the relativistic
corrections that contain the curvature perturbation $\varphi$ 
appear only at the third order in the equations
of motion.  
From below, we refer to the ``Newtonian dynamical equations'' as \eqref{eq:Newt1} and \eqref{eq:Newt2} without $\varphi$ terms, with the identification $\delta \to \delta_N$ and $\vvec \to \vvec_N$.
Note, however, that
in the Newtonian dynamics, there is no constraint 
beyond the equation of motion such that the initial condition~$\SD(\xvec)$
in Section~\ref{ssec:EdSmain}
is rather unconstrained, as opposed to the case in the relativistic dynamics
due to the full ADM energy constraint equation~(\ref{eq:conv}).
Finally, the linear-order ADM energy constraint equation 
\eqref{eq:admlin} can be written as, using the ADM momentum constraint~\eqref{eq:momentum} to linear order to replace $\kappa$ with $\chi$,
\beeq
\frac32H^2\Omega_m\delta=-\frac{\Delta}{a^2} \left( \varphi - H\chi \right) \equiv -\frac{\Delta}{a^2}\varphi_\chi \, ,
\eneq
indicating that we may identify the Newtonian potential $\Phi_N=-\CP_\chi$,
where $\CP_\chi$ is the linear-order curvature potential in the
zero shear gauge.

\subsection{Standard perturbation theory}
\label{ssec:spt}

Here we briefly summarize
the key equations for deriving the standard Fourier kernels and their
recurrence relations. A comprehensive  review on this topic can be found
in \cite{BECOET02} (and references therein).

By assuming the separability of the time and the spatial dependences,
the standard perturbation theory (SPT) takes a perturbative approach
to the nonlinear solution:
\begin{align}
\label{eq:form}
\delta_N(t,\kvec)&\equiv \sum_{n=1}^\infty D^n(t)
\left[\prod_i^n\int{d^3\qvec_i\over(2\pi)^3}~\hat\delta(\qvec_i)\right]
(2\pi)^3\delta^D(\kvec-\qvec_{12\cdots n})F_n^{(s)}(\qvec_1, \cdots,\qvec_n)
\equiv\sum_{n=1}^\infty D^n(t)\delta^{(n)}(\kvec)~,
\\
{\theta_N(t,\kvec)\over Hf_1}&\equiv \sum_{n=1}^\infty D^n(t)
\left[\prod_i^n\int{d^3\qvec_i\over(2\pi)^3}~\hat\delta(\qvec_i)\right]
(2\pi)^3 \delta^D(\kvec-\qvec_{12\cdots n})G_n^{(s)}(\qvec_1, \cdots,\qvec_n)
\equiv\sum_{n=1}^\infty D^n\theta^{(n)}(\kvec)~,
\end{align}
where $\delta^D$ is the Dirac delta function,
$\qvec_{12\cdots n}\equiv \qvec_1+\cdots+\qvec_n$, 
$\delta^{(n)}(\kvec)$ and $\theta^{(n)}(\kvec)$ are time-independent
$n$-th order perturbations, $F_n^{(s)}$ and
$G_n^{(s)}$ are the SPT kernels symmetrized over its arguments.
The (dimensionless) Newtonian linear-order growth factor~$D(t) \equiv D_1(t)/D_1(t_0)$ is normalized 
to unity at some early epoch~$t_0$ when the nonlinearities are ignored 
and satisfies the differential equation $\ddot D+2H\dot D-4\pi G\bar\rho_mD=0$.
The initial linear density perturbation is set up in terms of which the perturbative expansion is given,
$\delta_N(t_0,\kvec)\equiv\delta_1^{(1)}(t_0,\kvec)\equiv \hat\delta(\kvec)$.
With these decompositions in the Fourier space,
the LHS of the Newtonian dynamical equations become
\begin{equation}
\begin{split}
\dot\delta_N + \theta_N &= Hf_1\sum_{n=1}^\infty D^n
\left( n\delta^{(n)} -\theta^{(n)} \right)~,
\\
\dot\theta_N+2H\theta_N-4\pi G\bar\rho_m\delta_N
&= H^2f^2_1\sum{D^n\over2}\left[(1+2n)\theta^{(n)}-3\delta^{(n)} \right]~,
\end{split}
\end{equation}
where we adopted the usual assumption $\Omega_m = f_1 = 1$ 
in SPT and utilized
the relation between the growth factor and the growth rate
$\dot D=HDf_1$.
The RHS of the Newtonian dynamical equations are the convolution 
in the Fourier space:
\begin{equation}
\begin{split}
\left[ -{1\over a}\nabla\cdot(\delta_N\vvec_N) \right](\kvec) & = \int{d^3\QVEC_1\over(2\pi)^3}\int{d^3\QVEC_2\over(2\pi)^3}
(2\pi)^3\delta^D(\kvec-\QVEC_{12})
\alpha_{12}\theta_N(\QVEC_1,t)\delta_N(\QVEC_2,t)
\equiv Hf_1\sum_{n=1}^\infty D^n A_n(\kvec)~,
\\
\left\{ {1\over a^2}\nabla\cdot[(\vvec_N\cdot\nabla)\vvec_N] \right\}(\kvec) & = \int{d^3\QVEC_1\over(2\pi)^3}\int{d^3\QVEC_2\over(2\pi)^3}
(2\pi)^3\delta^D(\kvec-\QVEC_{12})
\beta_{12}\theta_N(\QVEC_1,t)\theta_N(\QVEC_2,t)
\equiv H^2f^2_1\sum_{n=1}^\infty D^n B_n(\kvec)~,
\end{split}
\end{equation}
where the vertex functions are defined as
\begin{equation}
\alpha_{12} \equiv
\alpha(\QVEC_1,\QVEC_2)\equiv1+{\QVEC_1\cdot\QVEC_2\over Q_1^2}
\qquad \text{and} \qquad
\beta_{12} \equiv\beta(\QVEC_1,\QVEC_2)\equiv
{|\QVEC_1+\QVEC_2|^2\QVEC_1\cdot\QVEC_2\over2 Q_1^2Q_2^2}~,
\end{equation}
and the $n$-th order perturbation kernels $A_n(\kvec)$ and $B_n(\kvec)$ are
\begin{equation}
\label{eq:Newtoniankernels}
\begin{split}
A_n(\kvec)&=
\left[\prod_i^n\int{d^3\qvec_i\over(2\pi)^3}
~\hat\delta(\qvec_i)\right](2\pi)^3\delta^D(\kvec-\qvec_{12\cdots n})
\sum_{i=1}^{n-1}\alpha_{12}
G_i(\qvec_1,\cdots,\qvec_i)F_{n-i}(\qvec_{i+1},\cdots,\qvec_n)~,
\\
B_n(\kvec)&=\left[\prod_i^n\int{d^3\qvec_i\over(2\pi)^3}
~\hat\delta(\qvec_i)\right](2\pi)^3\delta^D(\kvec-\qvec_{12\cdots n})
\sum_{i=1}^{n-1}\beta_{12}
G_i(\qvec_1,\cdots,\qvec_i)G_{n-i}(\qvec_{i+1},\cdots,\qvec_n)~,
\end{split}
\end{equation}
with $\QVEC_1=\qvec_{1\cdots i}$ and $\QVEC_1+\QVEC_2=\kvec$.

Therefore, the two Newtonian dynamical equations become algebraic equations
without time-dependence:
\begin{equation}
\label{eq:alge}
n\delta^{(n)} -\theta^{(n)}  = A_n~,\qquad\qquad
(1+2n)\theta^{(n)}-3\delta^{(n)} =2 B_n~,
\end{equation}
and the well-known recurrence formulas for the solutions are
\begin{equation}
\delta^{(n)} = {(1+2n) A_n+2 B_n\over(2n+3)(n-1)}
\qquad \text{and} \qquad 
\theta^{(n)} = {3 A_n+2n B_n\over(2n+3)(n-1)}~,
\end{equation}
and similarly so for the SPT kernels
\begin{equation}
\label{eq:recur}
\begin{split}
F_n&= \sum_{i=1}^{n-1}{G_i\over(2n+3)(n-1)}\left[(1+2n)\alpha_{12}F_{n-i}
+2\beta_{12}~G_{n-i}\right]~,
\\
G_n&= \sum_{i=1}^{n-1}{G_i\over(2n+3)(n-1)}
\left[3\alpha_{12}F_{n-i}+2n\beta_{12}G_{n-i}\right]~,
\end{split}
\end{equation}
with $F_1=G_1=1$. Using the recurrence relations~\eqref{eq:recur}, the SPT kernels
$F_n\sim G_n\propto k^2$ for $n>1$ in the limit $k\RA0$, with the individual
momentum~$\qvec_i$ held finite. This originates from the momentum
conservation of the nonlinear evolution.

\subsection{Relativistic effects in the density and velocity fluctuations}
\label{ssec:grkernel}

As emphasized,
the relativistic dynamical equations~\eqref{eq:dyn1} and~\eqref{eq:dyn2}
are identical to the standard Newtonian equations up to the second order
terms [see also~\eqref{eq:Newt1} and~\eqref{eq:Newt2}], 
and the relativistic terms ($\sim\CP$) appear only in the
third order terms in the RHS of the dynamical equations.
the Fourier decomposition of~$\delta_N$ and~$\theta_N$
in Section~\ref{ssec:spt} is valid for~$\delta$ and~$\kappa$, but the relativistic
corrections need to be further supplemented to the standard Newtonian solutions.
Since the curvature perturbation~$\CP$ is time-independent at the linear
order, the time-dependences of these relativistic corrections 
in the RHS of the dynamical equations are
\beeq
{1\over a^2}\chi\delta\CP\sim\dot DD~\hat\delta^2~\CP
\sim Hf_1D^2~\hat\delta^3~,
\qquad\qquad
{1\over a^2}\chi\kappa\CP\sim\dot D^2\hat\delta^2\CP\sim H^2f_1^2D^2\hat\delta^3~,
\eneq
and it is apparent that these terms will affect $\tilde\delta_2(\kvec)$
and $\tilde\kappa_2(\kvec)$ due to their time-dependence,
despite being at the third order in perturbations.
Note that the quadratic terms in the dynamical equations yield the 
standard $F_2$ and $G_2$ in Fourier space or~\eqref{eq:newtsol2}
and~\eqref{eq:newtsol2k} in configuration space.

To implement this change in the perturbative approach in~\eqref{eq:form},
we introduce a {\em time-dependent} third-order SPT kernel
\beeq
F_3^{\delta_2}(t,\kvec)\equiv{1\over D_1(t)}\tilde 
F_3^{\delta_2}(\kvec)~,\eneq
such that the density fluctuation is
\beeq
\delta(t,\kvec)\propto D\hat\delta+D^2F_2\hat\delta^2
+D^3\left(F_3+F_3^{\delta_2}\right)\hat\delta^3
=D\hat\delta+D^2\left(F_2-\tilde F_3^{\delta_2}
\Delta\II\right)\hat\delta^2+D^3F_3\hat\delta^3~,
\eneq
and similarly so for~$\kappa$ and~$G_3^{\kappa_2}$.
Therefore, the additional terms in the
algebraic equations~\eqref{eq:alge} are
\begin{equation}
2\tilde F_3^{\delta_2}-\tilde G_3^{\kappa_2}  =\tilde\CCC_1(\kvec)\equiv
{\CCC_1(t,\kvec)\over Hf_1D^2}
\qquad \text{and} \qquad
5\tilde G_3^{\kappa_2}-3\tilde F_3^{\delta_2} =
2\tilde\CCC_2(\kvec)\equiv{2\CCC_2(t,\kvec)\over H^2f_1^2D^2}~,
\end{equation}
where $\CCC_1$ and $\CCC_2$ represent respectively the third order terms in the relativistic
dynamical equations~\eqref{eq:Newt1} and~\eqref{eq:Newt2}.
The relativistic corrections to the SPT kernels are then
\begin{equation}
\tilde F_3^{\delta_2} ={5\tilde\CCC_1+2\tilde\CCC_2\over7}
\qquad \text{and} \qquad
\tilde G_3^{\kappa_2} ={3\tilde\CCC_1+4\tilde\CCC_2\over7}~.
\end{equation}
Using linear-order perturbation variables
\beeq
\kappa_1^{(1)}(t,\kvec)=Hf_1D\hat\delta(\kvec)={k^2\over a^2}~\chi_1^{(1)}
(t,\kvec)~,\qquad
\vvec^{(1)}(t,\kvec)=ia{\kvec\over k^2}\kappa_1^{(1)}(t,\kvec)~,\qquad
\CP(\kvec)={1\over k^2}\hat\delta(\kvec)~,
\eneq
the third order terms of the relativistic corrections 
can be computed \cite{NOHW08,JEGOET11} as
\begin{align}
\tilde\CCC_1&=
-{2\qvec_2\cdot\qvec_3\over q_1^2q_2^2}
-{\qvec_{12}\cdot\qvec_3\over q_{12}^2}
\left(-{2\over q_1^2}+{\qvec_1\cdot\qvec_2\over q_1^2q_2^2}-{3\over2}
{\qvec_{12}\cdot\qvec_2\over q_{12}^2q_2^2}
-{3\over2}{\qvec_{12}\cdot\qvec_1\over q_{12}^2}
{\qvec_1\cdot\qvec_2\over q_1^2q_2^2}\right)~,
\\
\tilde\CCC_2&=
{2\over3}{\qvec_2\cdot\qvec_3\over q_1^2q_2^2}
-4\left[{\kvec\cdot\qvec_3\over q_3^2}{\qvec_2\cdot\qvec_3\over  q_1^2q_2^2}
-{1\over3}{\kvec\cdot\qvec_2\over  q_1^2q_2^2}\right] 
\nonumber\\
& \quad
+\left[{2\over3}
+{\qvec_{12}\cdot\qvec_3\over q_3^2}\left(1-{k^2\over q_{12}^2}\right)
\right]
\left[-{2\over q_1^2}+{\qvec_1\cdot\qvec_2\over q_1^2q_2^2}
-{3\over2}{\qvec_{12}\cdot\qvec_2\over q_{12}^2q_2^2}
-{3\over2}{\qvec_{12}\cdot\qvec_1\over q_{12}^2q_1^2}
{\qvec_1\cdot\qvec_2\over q_2^2}\right] ~, 
\end{align}
where the kernels need to be symmetrized over the arguments.
Note that $F_3^{\delta_2}\propto1/(D_1k^2)\propto(\HH/k)^2$
is dimensionless, as expected.

This derivation of the relativistic corrections, so far, 
is essentially equivalent to those in \cite{NOHW08,JEGOET11}.
However, as we discussed in Section~\ref{ssec:compour}, the density fluctuation
at the early time~$t_0$ is {\it not} linear order 
due to the nonlinearity in the constraint equation~\eqref{eq:conv},
even if the initial condition~$\II$ is a linear-order Gaussian variable
and the initial epoch is set $t_0\to0$. To accommodate this intrinsic
nonlinearity to the standard Fourier kernels, we need to introduce
additional time-dependent kernels:
\begin{equation}
F_2^{\delta_1}(t,\kvec) \equiv{1\over D_1(t)}\tilde F_2^{\delta_1}(\kvec)
\qquad \text{and} \qquad
F_3^{\delta_1}(t,\kvec) \equiv{1\over D_1^2(t)}\tilde F_3^{\delta_1}
(\kvec)~,
\end{equation}
where two time-independent kernels are
\begin{equation}
\tilde F_2^{\delta_1} =-
{1\over k^2}\left[\frac32{k^2\qvec_1\cdot\qvec_2\over q_1^2q_2^2}
+2\left({k^2\over q_1^2}+{k^2\over q_2^2}\right)\right]
\qquad \text{and} \qquad
\tilde F_3^{\delta_1} =
3\left({\qvec_2\cdot\qvec_3\over q_1^2q_2^2q_3^2}+\up{cycl.}\right)
+4\left({1\over q_1^2q_2^2}+\up{cycl.}\right)~.
\end{equation}
As in~\eqref{eq:kkk1}, the higher-order terms in~$\kappa_1$ are identical 
to~$\delta_1$, and so are their kernels.

Similarly for~$\delta_2$ and~$\kappa_2$, this intrinsic nonlinearity of
the second order terms in~$F_2$ and~$G_2$ in the Fourier space or
\eqref{eq:newtsol2} and~\eqref{eq:newtsol2k} yields additional third-order
terms described in the last lines of~\eqref{eq:relsol3d} 
and~\eqref{eq:relsol3}, and this will modify $F_3^{\delta_2}$ and 
$G_3^{\kappa_2}$:
\begin{equation}
\Delta F_3^{\delta_2}(t,\kvec) \equiv{1\over D_1(t)}\Delta \tilde
F_3^{\delta_2}(\kvec)
\qquad \text{and} \qquad
\Delta G_3^{\kappa_2}(t,\kvec) \equiv{1\over D_1(t)}\Delta \tilde
G_3^{\kappa_2}(\kvec)~,
\end{equation}
where two time-independent spatial kernels are
\begin{equation}
\begin{split}
\Delta\tilde F_3^{\delta_2}(\kvec)&={2^2\over5^2}
\left[{10\over7}+{\qvec_1\cdot\qvec_{23}\over q_1^2q_{23}^2}
\left(q_1^2+q_{23}^2\right)
+\frac47{(\qvec_1\cdot\qvec_{23})^2\over q_1^2q_{23}^2}
\right]\tilde F_2^{\delta_1}(\qvec_2,\qvec_3)~,
\\
\Delta\tilde G_3^{\kappa_2}(\kvec)&={2^2\over5^2}
\left[{6\over7}+{\qvec_1\cdot\qvec_{23}\over q_1^2q_{23}^2}
\left(q_1^2+q_{23}^2\right)
+\frac87{(\qvec_1\cdot\qvec_{23})^2\over q_1^2q_{23}^2}
\right]\tilde F_2^{\delta_1}(\qvec_2,\qvec_3)~,
\end{split}
\end{equation}
where the kernels need to be symmetrized over the arguments.

\section{Discussions}
\label{sec:dis}

The proper-time hypersurface of nonrelativistic matter flows is a physically
well-defined global
 time-slicing that a local observer moving with nonrelativistic 
matter can establish. Galaxy bias in the Newtonian context can be naturally
generalized in this proper-time hypersurface in the relativistic context
\cite{YOO14b}. As the first step toward this direction, we have derived
the third-order analytic solutions for the matter density and the velocity
fluctuations in the proper-time hypersurface, providing essential ingredients
for computing the subtle one-loop corrections to the matter power spectrum.

For the first time, we have derived the {\em exact} analytic solutions 
of the matter density and the velocity fluctuations in 
a $\Lambda$CDM universe, accounting for the nonlinear relativistic effects and
greatly extending the results of \cite{YOGO15} in 
the EdS universe.
Our general approach to solving the nonlinear dynamical equations allows 
us to derive the solutions in a $\Lambda$CDM universe, in which the 
time-dependence of
the solutions is more complicated than that in the EdS universe.
In particular, we have derived the explicit solutions to the Green's functions
for the growth factors, for which only the differential equations 
were known in literature.
Our solutions are composed of the standard Newtonian solutions and
the relativistic corrections. Our Newtonian solutions with the exact
time-dependences show that the standard assumption that
the solution is separable in its time and spatial dependences is {\it invalid},
rendering the growth of perturbations scale-dependent in general relativity.
However, the recent study in \cite{TAKA08} of the Newtonian perturbation theory
shows that as long as the linear-order growth factor is properly considered,
the errors in the power spectrum with the standard assumption 
are rather small at $k<0.2\hmpci$, while it amounts 
to $\sim0.5-1\%$ at $k\gtrsim0.2\hmpci$.

On large scales, which is the scale of our interest, the relativistic 
effects in galaxy clustering become important, providing unique opportunities 
to probe subtle properties of gravity and the physics relevant for 
the early universe. For example, the primordial non-Gaussianity can be
probed with the galaxy power spectrum via its unique scale-dependence on 
large scales \cite{DADOET08}. As this unique signature of the
early universe is also a relativistic effect, we need to take into 
consideration other relativistic effects in measuring the primordial
non-Gaussianity signature \cite{YOHAET12}.
The matter density fluctuation constitutes the dominant contribution to the
galaxy clustering measurements on all scales, and we have derived 
the exact relativistic corrections in a $\Lambda$CDM universe to the 
matter density fluctuation.
Previously, the third-order relativistic solutions for the matter density
and the velocity fluctuations were derived \cite{JEGOET11} in the
comoving gauge, assuming the EdS universe. For a presureless
medium, the comoving gauge condition corresponds to the proper-time 
hypersurface \cite{YOO14b}. Their solutions agree with ours in the Newtonian
part, while there exist differences in the relativistic corrections.
The nonlinear constraint equations in general relativity impose
nonlinearity in the matter density fluctuation at early time, 
even with the initial condition set up by the comoving-gauge curvature
potential at the linear order in perturbations. We have demonstrated that
the difference in the two solutions is exactly due to the initial nonlinearity
in the matter density fluctuation, imposed by the ADM energy constraint. 
Given that the initial condition is set up by inflation
at early time, when there is no matter fluid to begin with, our solutions are 
more appropriate for analyzing the nonlinear growth of the matter density
fluctuation in general relativity.

In the era of precision measurements from numerous current and future
galaxy surveys, the subtle relativistic effects in galaxy clustering
can be utilized to distinguish various inflationary models or competing
dark energy models on large scales. The third-order analytic solutions
for the matter density fluctuation in this work provide such a first step.

\subsection*{Acknowledgments}

J.~G. is grateful to the Center for Theoretical Astrophysics and Cosmology, Universit\"at Z\"urich
for hospitality while this work was finalized.
J.~Y. is supported by the Swiss National Science Foundation and
a Consolidator Grant of the European Research Council (ERC-2015-CoG grant
680886).
J.~G. acknowledges support from the Korea Ministry of Education, 
Science and Technology, Gyeongsangbuk-Do and Pohang City for 
Independent Junior Research Groups at the Asia Pacific Center for 
Theoretical Physics. 
J.~G is also supported in part by a Starting Grant through 
the Basic Science Research Program of the National Research Foundation of Korea (2013R1A1A1006701) 
and by a TJ Park Science Fellowship of POSCO TJ Park Foundation.

\end{document}